\documentclass[%
 reprint,
superscriptaddress,
 amsmath,amssymb,
 aps,
 pra,
]{revtex4-2}

\usepackage{graphicx}
\usepackage{dcolumn}
\usepackage{bm}
\usepackage[hidelinks]{hyperref}
\usepackage{tikz}
\usepackage{pifont}

\usepackage{acro}
\usepackage{physics}
\usepackage{mathtools}

\usepackage{enumitem}

\newcommand{\vect}[1]{\bm{#1}}

\newcommand{\fbinary}{\mathbb{F}_2}
\newcommand{\freal}{\mathbb{R}}

\newcommand{\iZ}{\mathrm{Z}}
\newcommand{\iX}{\mathrm{X}}
\newcommand{\iY}{\mathrm{Y}}
\newcommand{\iI}{\mathrm{I}}

\usepackage{tikz}
\usetikzlibrary{matrix,fit}

\usepackage{colortbl}

\newcommand{\llr}{L}

\usepackage[caption=false]{subfig}

\DeclareAcronym{CSS}{
  short = CSS ,
  long  = Calderbank-Shor-Steane
}

\DeclareAcronym{LDPC}{
  short = LDPC,
  long  = low-density parity-check
}

\DeclareAcronym{QLDPC}{
  short = QLDPC,
  long  = quantum LDPC
}

\DeclareAcronym{BP}{
  short = BP,
  long  = belief propagation
}

\DeclareAcronym{OSD}{
  short = OSD,
  long  = ordered statistics decoding
}

\DeclareAcronym{SOGRAND}{
  short = SOGRAND,
  long  = soft-output GRAND
}

\DeclareAcronym{GRAND}{
  short = GRAND,
  long  = guessing random additive noise decoding
}

\DeclareAcronym{ORBGRAND}{
  short = ORBGRAND,
  long  = ordered reliability bits guessing random additive noise decoding
}

\DeclareAcronym{APP}{
  short = APP,
  long = a posteriori probability
}

\DeclareAcronym{SISO}{
  short = SISO,
  long  = soft-input soft-output
}

\DeclareAcronym{BLER}{
  short = BLER,
  long  = block error rate
}

\DeclareAcronym{LLR}{
  short = LLR,
  long  = log-likelihood ratio
}

\DeclareAcronym{GLDPC}{
  short = GLDPC,
  long  = generalized LDPC,
  first-style = long,   
  subsequent-style = long,   
}

\DeclareAcronym{VN}{
  short = VN,
  long  = variable node,
  long-plural-form = variable nodes,
  first-style = long,   
  subsequent-style = long,    
}

\DeclareAcronym{CN}{
  short = CN,
  long  = constraint node,
  long-plural-form = constraint nodes,
  first-style = long,   
  subsequent-style = long,    
}

\begin{document}

\title{Efficient Soft-Output Guessing for Enhanced Quantum Tanner Code Decoding}

\author{Lukas Rapp}
\email{rappl@mit.edu}
\affiliation{Research Laboratory for Electronics, Massachusetts Institute of Technology, Cambridge, MA, USA}

\author{Muriel Médard}
\email{medard@mit.edu}
\affiliation{Research Laboratory for Electronics, Massachusetts Institute of Technology, Cambridge, MA, USA}

\author{Eugene Tang}
\email{e.tang@northeastern.edu}
\affiliation{Department of Mathematics \& Physics, Northeastern University, Boston, MA, USA}

\author{Ken R. Duffy}
\email{k.duffy@northeastern.edu}
\affiliation{Department of Mathematics \& ECE, Northeastern University, Boston, MA, USA}

\date{\today}

\begin{abstract} 
    We introduce a generalized low-density parity-check decoding framework for quantum Tanner codes utilizing soft-output guessing random additive noise decoding (SOGRAND). By soft-output decoding entire component codes rather than individual parity checks, we mitigate the effects of trapping sets and cycles, resulting in improved convergence. Because our decoder preserves the message passing structure of standard belief propagation (BP), it is compatible with many BP enhancements. We demonstrate this with OSD postprocessing, quaternary BP, and RelayBP, each yielding further gains. 
    Standalone SOGRAND outperforms the standard BP+OSD baseline by over two orders of magnitude in logical error rate. In combination with the Relay principle, SOGRAND outperforms RelayBP by over one order of magnitude, providing a way forward for scalable decoding of the emerging class of Tanner-code-based quantum codes.
\end{abstract}

\maketitle

\section{Introduction}
Quantum error correction is an essential component of large-scale quantum computing. Although the surface code~\cite{Kitaev_2003,bravyi1998quantumcodeslatticeboundary} has long been the dominant approach to fault tolerance, high-rate quantum low-density parity-check (QLDPC) codes have recently attracted significant attention due to their potential to greatly reduce qubit overhead~\cite{MacKay_2004,tillichQuantumLDPCCodes2014,Breuckmann_2021,gottesmanFaulttolerantQuantumComputation2014,Breuckmann_2021b}. The discovery of asymptotically good QLDPC codes~\cite{panteleevAsymptoticallyGoodQuantum2022}, and the subsequent development of quantum Tanner codes~\cite{leverrierQuantumTannerCodes2022}, marked a major milestone in efficient quantum error correction. The theoretical promise of quantum Tanner codes has fueled growing interest in finite-length instances suitable for near-term implementation~\cite{leverrierSmallQuantumTanner2025, guemardModeratelengthLiftedQuantum2025, radeboldExplicitInstancesQuantum2025}, but key obstacles remain, foremost among them the problem of efficient decoding.

Practical implementations require real-time decoding that keeps up with the syndrome generation to avoid an exponential data backlog~\cite{terhalQuantumErrorCorrection2015}. To date, \ac{QLDPC} codes have primarily been decoded using \ac{BP}, the standard decoder for classical \ac{LDPC} codes~\cite{gallagerLowDensity1963}. However, \ac{BP} struggles in the quantum regime, as stabilizer commutativity enforces short cycles, and high degeneracy leads to trapping sets, preventing convergence to a syndrome-consistent error pattern~\cite{babarFifteenYearsQuantum2015a, poulinIterativeDecodingSparse2008, raveendranTrappingSetsQuantum2021}. Fundamentally, \ac{BP}'s ability to decode accurately is limited because it relies on the local soft-output decoding of weak single-parity-check constraints. 

Current proposals include postprocessing, such as \ac{OSD}~\cite{fossorierSoftdecisionDecodingLinear1995add, panteleevDegenerateQuantumLDPC2021}, applied whenever \ac{BP} fails (\ac{BP}+\ac{OSD}). While this approach improves decoding performance, it does so at a computational complexity that is at least cubic in the code length. The OSD postprocessing step is generally not considered feasible for near-term VLSI implementation~\cite{vallsSyndromeBasedMinSumVs2021}. However, promising implementation results have been reported for specific code families, including surface codes and bivariate bicycle codes~\cite{Bascones2025FPGAASICBPOSD}. Recently, localized statistics decoding~\cite{hillmannLocalizedStatisticsDecoding2025add}, a faster OSD alternative, and \ac{BP} with guided decimation~\cite{yaoBeliefPropagationDecoding2024aadd, gongLowlatencyIterativeDecoding2024add} have been proposed. New search-based decoders~\cite{beniTesseractSearchBasedDecoder2025, ottDecisiontreeDecodersGeneral} improve upon BP+OSD in performance and decoding speed. However, it remains unclear whether they admit compact hardware implementations~\cite{mullerImprovedBeliefPropagation2025}.

Another promising direction is to enhance BP's message-passing procedure. Examples include neural belief propagation~\cite{liuNeuralBeliefPropagationDecoders2019a, miaoQuaternaryNeuralBelief2025add} and efficient quaternary \ac{BP} with message normalization~\cite{laiLogDomainDecodingQuantum2021a}, as well as several techniques that introduce memory effects in \ac{BP}'s variable node update, such as \ac{BP} with additional memory effects (MBP)~\cite{kuoExploitingDegeneracyBelief2022add}, EWAInit-BP~\cite{chenImprovedBeliefPropagation2025add}, MS-PI~\cite{chytasEnhancedMinSumDecoding2025a} and RelayBP~\cite{mullerImprovedBeliefPropagation2025, maurerRealtimeDecodingGross2025}. RelayBP performs well across various \ac{QLDPC} codes, particularly bivariate bicycle codes~\cite{bravyiHighthresholdLowoverheadFaulttolerant2024a, yoderTourGrossModular2025, shirizlyFeedforwardSuppressionReadoutinduced2025a}.

However, none of these techniques exploit the graphical structure specific to quantum Tanner codes. In this paper, we offer an alternative way forward by exploiting that graphical structure in conjunction with recent developments in classical soft output decoding. Rather than decoding individual constraints, we treat the code as a \emph{generalized} \ac{LDPC} code~\cite{boutrosGeneralizedLowDensity1999} and iteratively soft-decode its more powerful component codes. Unlike previous studies with bit-flip decoding that focus on identifying asymptotic bounds~\cite{Dinur2023,Gu2023,leverrierDecodingQuantumTanner2022, guSingleShotDecodingGood2024a, leverrierEfficientDecodingConstant2025}, this proposal processes soft messages, a strategy known to outperform bit-flipping in classical \ac{LDPC} codes~\cite{gallagerLowDensity1963}. By replacing single-parity-check updates with joint decoding of multiple constraints, short cycles and trapping sets are mitigated. 

A central challenge in implementing a \ac{GLDPC} framework for quantum codes is the requirement of an efficient \ac{SISO} decoder for the component codes. Until recently, there were few practical approaches to \ac{SISO} decoding of multi-constraint component codes. Most of them require hyperparameters to manage the inability to estimate the likelihood that the true error pattern has not been found.
For example, Ref.~\cite{pyndiahNearoptimumDecodingProduct1998added} computes each \ac{LLR} from the most likely codeword and the competing codeword that differs at that position. If no competing codeword is found,  a predefined, iteration-dependent reliability is used instead.
This has changed recently with the establishment of \ac{SOGRAND}~\cite{yuanSoftoutputGRANDLong2023}, developed from the \ac{GRAND} methodology~\cite{duffyCapacityAchievingGuessingRandom2019}.
\ac{SOGRAND} is able to provide accurate \ac{SISO} decoding for any moderate-redundancy component code and has been demonstrated in a taped-out chip~\cite{Kizilates26SOGRAND}, which is why we adopt it for use in this new setting.

Recently, the original hard-input, hard-output \ac{GRAND} framework has been extended to the quantum regime, demonstrating potential for hard-decision decoding of high-rate quantum codes~\cite{cruzQuantumErrorCorrection2023, chandraUniversalDecodingQuantum2023, cruzFaultTolerantNoiseGuessing2025}. While hard-decision decoding operates without reliability information, we adapt \ac{SOGRAND}, a more recent and sophisticated \ac{SISO} version of \ac{GRAND} that enhances iterative decoding performance. By \ac{SISO} decoding local component codes, this technique overcomes the rate limitations inherent in previous \ac{GRAND}-based approaches. Although originally designed for classical channels with heterogeneous soft reliability as input, we demonstrate its effectiveness with quantum codes and the statistical soft input inherent to the quantum setting.

The proposed standalone decoder is demonstrated to outperform the \ac{BP}+\ac{OSD} baseline by over two orders of magnitude in logical error rate on instances of quantum Tanner codes~\cite{radeboldExplicitInstancesQuantum2025} while significantly accelerating convergence.
Because \ac{SOGRAND} decoding relies on iterative message passing and preserves the \ac{VN} of standard \ac{BP}, many existing postprocessing techniques and \ac{BP} enhancements can be readily adopted. In this paper, we demonstrate this with \ac{OSD} postprocessing~\cite{panteleevDegenerateQuantumLDPC2021}, quaternary \ac{BP}~\cite{laiLogDomainDecodingQuantum2021a}, and the Relay principle proposed for RelayBP~\cite{mullerImprovedBeliefPropagation2025}, each yielding additional performance improvements. Combined with the Relay principle, \ac{SOGRAND} outperforms RelayBP by over one order of magnitude in logical error rate.
As Tanner code-based constructions are central to the recent development of asymptotically good \ac{QLDPC} codes~\cite{dinurLocallyTestableCodes2022, panteleevAsymptoticallyGoodQuantum2022, leverrierQuantumTannerCodes2022, mostadAsymptoticallyGoodGeneralized2025}, this approach serves as a powerful alternative foundation for efficient, real-time decoding across a broad class of emerging quantum error-correcting codes. 

\acuse{SOGRAND}
\acuse{GRAND}

\section{Preliminaries}

\subsection{Stabilizers and noise}
We consider an \([[n, k, d]]\) stabilizer code~\cite{calderbankGoodQuantumErrorcorrecting1996, steaneErrorCorrectingCodes1996} defined as the joint \(+1\) eigenspace of an abelian subgroup \(\mathcal{S} \subset \mathcal{P}_n\), where \(\mathcal{P}_n\) denotes the \(n\)-qubit Pauli group and \(-I \notin \mathcal{S}\). As exemplars, we focus on Quantum Tanner codes, which are \ac{CSS} codes. A \ac{CSS} code is defined by two parity-check matrices \(H_\iX \in \fbinary^{m_\iX \times n}, H_\iZ \in \fbinary^{m_\iZ \times n}\), which specify the \(X\) and \(Z\) stabilizer generators, respectively and satisfy the \ac{CSS} condition: \(H_\iX H_\iZ^T = 0\). 
While our analysis focuses on \ac{CSS} codes, the proposed decoder naturally extends to any quantum code that admits a suitable \ac{GLDPC} structure.
An error 
\[
    E = \bigotimes_{i=1}^n X^{e_{\iX,i}} Z^{e_{\iZ,i}} \in \mathcal{P}_n
\] 
is represented by a binary vector \(\vect{e} = (\vect{e_\iX} | \vect{e_\iZ}) \in \fbinary^{2n}\), which yields syndrome pairs 
\[
    \vect{s_\iX} = H_\iZ \vect{e_\iX}, \qquad \vect{s_\iZ} = H_\iX \vect{e_\iZ}.
\]
The decoder uses \(\vect{s_\iX}\), and \(\vect{s_\iZ}\) to estimate the error patterns \(\vect{\hat{e}} = (\vect{\hat{e}_\iX} | \vect{\hat{e}_\iZ})\). The correction succeeds if the residual \(\vect{e} \oplus \vect{\hat{e}}\) belongs to the stabilizer group \(\mathcal{S}\).

As a starting point, we consider the code capacity setting where only data qubits experience errors and syndrome measurements are assumed to be perfect. For the error model, we consider independent and identically distributed qubit errors \(E_i\). For pedagogical simplicity, we first describe decoding assuming independent \(X\) and \(Z\) errors, before incorporating \(X/Z\) correlation when discussing BP enhancements in Sec.~\ref{sec:impr}. Owing to the symmetry, we focus on the  \(Z\)-error decoder. The decoder is initialized with a vector of  \acp{LLR} \(\vect{L_\mathrm{ch}}\) that encapsulate a priori beliefs.
Given \(\vect{L_\mathrm{ch}}\), the decoder tries to find an error pattern \(\vect{\hat{e}_\iZ}\) that maximizes the posterior probability \(P(\vect{e_\iZ}=\vect{\hat{e}_\iZ} | \vect{s_\iZ})\) while satisfying the \emph{syndrome condition}: 
\begin{equation}\label{eqn:syndrome-condition}
    H_\iX \vect{\hat{e}_\iZ} = \vect{s_\iZ}.
\end{equation}

\subsection{Quantum Tanner codes}

A classical Tanner code~\cite{tannerRecursiveApproachLow1981} is defined by a linear binary component code, represented by a parity-check matrix \(H_\mathrm{c}\), and a bipartite graph. The bipartite graph consists of two node types with edges that only connect different node types. For the classical Tanner code, these are \(n\) \emph{\acp{VN}}, associated with the bits, and \(m\) \emph{\acp{CN}}, associated with local constraints. While the degree distribution of the graph can in general be arbitrary~\cite{tannerRecursiveApproachLow1981}, we restrict our attention to the case where the variable nodes are of degree $2$, as required by the quantum Tanner code construction (see Fig.~\ref{fig:overview} left side).
Each \ac{CN} \(j \in [m] = \{1, \dots, m\}\) enforces a local constraint from \(H_\mathrm{c}\) on its incident \acp{VN}. Specifically, let \(\mathcal{M}(j)\) be an ordering of the edges connected to \ac{CN} \(j\). The local view of a vector \(\vect{x}\) at \(j\) is the subvector \(\vect{x^j} = (x_i)_{i \in \mathcal{M}(j)}\), where the ordering is defined by \(\mathcal{M}(j)\). 
A string \(\vect{c} \in \fbinary^n\) is a codeword if and only if the local view at every \ac{CN} \(j \in [m]\) is a valid component codeword: \(H_\mathrm{c} \vect{c^j} = 0\). As a linear code, a Tanner code is fully characterized by a global parity-check matrix \(H\).

\emph{Quantum Tanner codes}~\cite{leverrierQuantumTannerCodes2022} are \ac{CSS} codes where \(H_{\iX}\) and \(H_{\iZ}\) are the global parity-check matrices of two classical Tanner codes with component parity-check matrices \(H_{\mathrm{Xc}}, H_{\mathrm{Zc}}\). They are constructed using an expanding left-right Cayley complex~\cite{dinurLocallyTestableCodes2022}, with component codes chosen as dual product codes to ensure that the \ac{CSS} condition is fulfilled.
Since quantum Tanner codes belong to the class of \ac{QLDPC} codes, i.e., \(H_\iX, H_\iZ\) are sparse, they can be decoded with \ac{BP}~\cite{gallagerLowDensity1963} as demonstrated in Ref.~\cite{radeboldExplicitInstancesQuantum2025}. In this framework, bit beliefs about error patterns are represented as \acp{LLR}. After initializing them with \(\vect{L_\mathrm{ch}}\), they are iteratively exchanged between variable and check nodes until Eq.~\eqref{eqn:syndrome-condition} is satisfied or a maximum number of decoding iterations \(N_\mathrm{iter}\) is reached.

\section{SOGRAND Decoding of Quantum Tanner Codes}\label{sec:sogrand-method}

\subsection{Decoding of Generalized LDPC Codes}
As quantum Tanner codes are constructed from classical Tanner codes, they are also compatible with iterative decoding techniques originally proposed for their classical counterparts~\cite{tannerRecursiveApproachLow1981,boutrosGeneralizedLowDensity1999}.
Instead of treating the Tanner codes as \ac{QLDPC} 
codes, we can treat them as \emph{generalized} \ac{LDPC} codes, leveraging the underlying graph structure, and decode the local views of the component codes rather than weak single-parity-check codes. This approach is made possible by recent advances in classical \ac{SISO} decoding of arbitrary multi-bit redundancy codes~\cite{yuanSoftoutputGRANDLong2023,duffySoftOutputGuessingCodeword2025a,yuanSoftOutputSuccessiveCancellation2025}. These techniques were initially applied to the decoding of classical product codes, which are concatenations of component codes, for which \ac{SISO} component decoding is an important strategy~\cite{pyndiahNearoptimumDecodingProduct1998added, yuanSoftoutputGRANDLong2023}.

\begin{figure}
    \centering
    \includegraphics{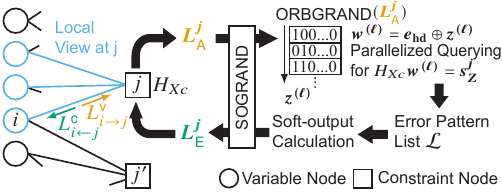}
    \caption{Overview of SOGRAND decoding on \ac{GLDPC} codes. Decoding is performed through iterative message passing on the classical Tanner code (left). During the \ac{CN} update (right), \ac{SOGRAND} decodes the incoming \acp{LLR} \(\vect{L_\mathrm{A}^j}\) and computes extrinsic \acp{LLR} \(\vect{L_\mathrm{E}^j}\). Within this process, \acs{SOGRAND} identifies a list of likely error patterns that satisfy the local syndrome condition, as well as a likelihood that the correct error pattern has not been identified, from which soft-output is extracted.}
    \label{fig:overview}
\end{figure}

To formalize \ac{BP} under this \ac{GLDPC} interpretation, let \(L_{i \rightarrow j}^{\mathrm{v}}, L_{i \leftarrow j}^{\mathrm{c}} \in \freal\) be the messages sent from \ac{VN} \(i \in [n]\) to \ac{CN} \(j \in [m_\iX]\) and from \ac{CN} \(j\) to \ac{VN} \(i\), respectively (Fig.~\ref{fig:overview} left). A message sent to or from a \ac{VN} \(i\) is an \ac{LLR}, representing a local belief about the \(i\)-th bit of the error pattern \(\vect{e_\iZ}\). In \ac{BP}, these messages are iteratively processed by node updates until a termination condition is fulfilled. For \ac{GLDPC} codes, these update rules consist of:

\begin{description}[leftmargin=0pt]
\item[Initialization] For each \ac{VN} \(i \in [n]\), both outgoing messages are initialized as \(L_{i \rightarrow j}^{\mathrm{v}} = L_{\mathrm{ch},i}\).
\item[\Ac{CN} update] In parallel, each \ac{CN} \(j \in [m_\iX]\) receives a message vector 
\(\vect{\llr_\mathrm{A}^j} \gets (L_{i \rightarrow j}^\mathrm{v})_{i \in \mathcal{M}(j)}\). A \ac{SISO} decoder takes \(\vect{\llr_\mathrm{A}^j}\) as soft-input, decodes the local view using the component code \(H_\mathrm{Xc}\) and outputs extrinsic information \(\vect{\llr_\mathrm{E}^j}\). These extrinsic \acp{LLR} are sent back to the connected \acp{VN}: \((L_{i \leftarrow j}^{\mathrm{c}})_{i \in \mathcal{M}(j)} \gets \vect{\llr_\mathrm{E}^j}\).
\item[\Ac{VN} update] Each variable node \(i \in [n]\) receives messages from its two connected \acp{CN} \(j, j^\prime \in \mathcal{N}(i)\). It calculates the \ac{APP} \acp{LLR} 
\begin{equation}\label{eqn:vn-app-calc}
    L_{\mathrm{APP},i} \leftarrow L_{\mathrm{ch},i} + L_{i \leftarrow j}^{\mathrm{c}} + L_{i \leftarrow j^\prime}^{\mathrm{c}}
\end{equation}
and hard-decision \(\hat{e}_{\iZ, i}\), which is \(\hat{e}_{\iZ, i} = 0\) if \(L_{\mathrm{APP},i} \geq 0\) and \(1\), otherwise. In this context, \ac{APP} refers to a \ac{LLR} \(L_{\mathrm{APP},i}\), obtained by updating the a priori \ac{LLR} \(L_{\mathrm{ch},i}\) with new evidence from the syndrome condition \eqref{eqn:syndrome-condition}.

Finally, it sends the extrinsic messages to both \acp{CN} \(j \in \mathcal{N}(i)\): \(L_{i \rightarrow j}^{\mathrm{v}} \gets L_{\mathrm{APP},i} - L_{i \leftarrow j}^{\mathrm{c}}\).
\end{description}

During each \ac{VN} update, the decoder checks whether the current hard-decision \(\vect{\hat{e}_\iZ}\) fulfills Eq.~\eqref{eqn:syndrome-condition}. If this is the case or the maximal number of iterations \(N_\mathrm{iter}\) is reached, decoding terminates and \(\vect{\hat{e}_\iZ}\) is output.

\subsection{Soft Component Decoding} \label{sec:soft-dec}
Consider a \ac{CN} \(j \in [m_\iX]\) with parity-check matrix \(H_\mathrm{Xc} \in \fbinary^{m_\mathrm{Xc} \times n_\mathrm{Xc}}\) that receives the soft-input \(\vect{\llr_\mathrm{A}^j} \in \fbinary^{n_\mathrm{Xc}}\). We define the \emph{local syndrome} 
\(
    \vect{s_\iZ^j} = H_\mathrm{Xc} \vect{e_\iZ^j}
\)
at \ac{CN} \(j\) as a subvector of the global syndrome \(\vect{s_\iZ}\) corresponding to the stabilizers induced by \(j\), where \(\vect{e_\iZ^j}\) is the local view of \(\vect{e_\iZ}\) at \(j\). To produce accurate soft-output, \ac{SOGRAND} identifies a list \(\mathcal{L}\) of candidate error patterns \(\vect{w} \in \fbinary^{n_\mathrm{Xc}}\) that satisfy the local syndrome constraint,
\begin{equation}\label{eqn:syndrome-condition2}
    H_\mathrm{Xc} \vect{w} = \vect{s_\iZ^j},
\end{equation}
where \(\vect{L_\text{A}^{j}}\) determines the a priori probability
\[
    P(\vect{w} | \vect{\llr_\mathrm{A}^{j}}) 
    =
    \prod_{i=1}^{n_\mathrm{Xc}}
    \left(1+\exp(-(-1)^{w_i} L_{A,i}^j)\right)^{-1},
\]
of each pattern \(\vect{w}\). 

\ac{SOGRAND} finds candidate error patterns via the \ac{GRAND} decoding principle~\cite{duffyCapacityAchievingGuessingRandom2019}. Prior to decoding, \ac{SOGRAND} identifies the most likely local error pattern \(\vect{e_\mathrm{hd}}\), called the hard-decision, where \(e_{\mathrm{hd},i}\) is \(0\) if \(L_{\text{A},i}^j > 0\) and \(1\) otherwise.
The core component of a \ac{GRAND} decoder is a pattern generator that generates, every time it is queried, a binary perturbation \(\vect{z^{(\ell)}}\) that perturbs the hard decision, generating an error pattern query \(\vect{w^{(\ell)}} = \vect{e_{\mathrm{hd}}} \oplus \vect{z^{(\ell)}}\), where \(\oplus\) denotes binary addition. The generator approximates an optimal guessing order, in which error patterns are queried in decreasing order of likelihood, i.e., 
\(
    P(\vect{w^{(\ell)}} | \vect{L_\text{A}^{j}}) 
    \geq 
    P(\vect{w^{(\ell+1)}} | \vect{L_\text{A}^{j}})
\) 
for all query indices \(\ell\).
Each pattern \(\vect{w^{(\ell)}}\) that satisfies the local syndrome condition in Eq.~\eqref{eqn:syndrome-condition2} is added to \(\mathcal{L}\) until a maximal list size \(\mathcal{L}_\mathrm{max}\) is reached or a termination condition is met.
Since SOGRAND queries error patterns in approximately decreasing order of likelihood, it ensures that the resulting list contains approximately the most probable syndrome-consistent error pattern. 
While pattern generators have been proposed for various channels~\cite{duffyCapacityAchievingGuessingRandom2019, duffyOrderedReliabilityBits2022, duffyUsingChannelCorrelation2023aprl}, we employ \ac{ORBGRAND}~\cite{duffyOrderedReliabilityBits2022} for its hardware efficiency~\cite{condoHighperformanceLowcomplexityError2021, condoFixedLatencyORBGRAND2022, abbasHighThroughputEnergyEfficientVLSI2022, riazSub08pJBitUniversal2025}. 

To extract soft-output from \(\mathcal{L}\), \ac{SOGRAND} keeps track of the total probability mass \(P_\text{g}\) of all explored error patterns and of the mass 
of the patterns in the list \(\mathcal{L}\), where
\[
    P_\text{g} = \sum_{\ell=1}^{N_\text{guess}} P(\vect{w^{(\ell)}} | \vect{\llr_\mathrm{A}^{j}}), \quad
    P_{\mathcal{L}} = \sum_{\vect{\hat{w}} \in \mathcal{L}} P(\vect{\hat{w}} | \vect{\llr_\mathrm{A}^{j}})
    ,
\]
and \(N_\text{guess}\) denotes the number of generated error patterns.
Given there are \(2^{n_\mathrm{Xc}-m_\mathrm{Xc}}\) patterns \(\vect{w^{(\ell)}}\) satisfying Eq.~\eqref{eqn:syndrome-condition2}, and assuming they are uniformly distributed within the generator's guessing order, the probability that a syndrome-consistent error pattern is not in \(\mathcal{L}\) can be estimated as \(P_{\mathcal{L}^c} = (1-P_{\mathrm{g}}) 2^{-m_\mathrm{Xc}}\)~\cite{yuanSoftoutputGRANDLong2023}.
The total syndrome-consistent probability mass \(P_\mathrm{tot} = P_{\mathcal{L}} + P_{\mathcal{L}^c}\) enables the estimation of the \acp{APP} conditioned on Eq.~\eqref{eqn:syndrome-condition2} with local syndrome \(\vect{s_\iZ^j}\), \emph{i.e.}, the probability that a syndrome-consistent error pattern satisfying Eq.~\eqref{eqn:syndrome-condition2} corresponds to the correct local error pattern. According to Bayes' rule, this probability is given by:
\[
    P(\vect{\hat{w}} | \vect{\llr_\mathrm{A}^{j}}, \text{Eq.}\eqref{eqn:syndrome-condition2})
    = P(\vect{\hat{w}} | \vect{\llr_\mathrm{A}^{j}}) / P_\text{tot}.
\]
for \(\vect{\hat{w}} \in \mathcal{L}\), which concentrates the probability mass on syndrome-consistent error patterns.

This framework enables accurate soft-output estimation by incorporating the uncertainty of the unexplored space when calculating bit-wise marginals \(\vect{\llr_{\mathrm{APP}}^{j}}\) (see Ref.~\cite{yuanSoftoutputGRANDLong2023} for details). Finally, extrinsic information is computed: \(\vect{\llr_\mathrm{E}^{j}} = \vect{\llr_{\mathrm{APP}}^{j}} - \vect{\llr_\mathrm{A}^{j}}\). Extrinsic messages exclude the a priori knowledge \(\vect{L_\text{A}^j}\) from the decoding output, extracting the newly learned information to avoid positive feedback loops~\cite{berrouShannonLimitErrorcorrecting1993add}.

\subsection{Example}\label{sec:example}
\begin{figure}
    \centering
    \includegraphics{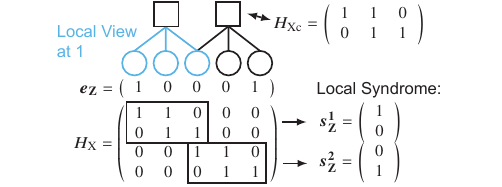}
    \caption{Example of a small-sized generalized LDPC code \(H_\iX\), consisting of two constraint nodes with component code \(H_\mathrm{Xc}\). We assume that the qubits are affected by the errors \(\vect{e_\iZ}\) which can be observed through the local syndromes \(\vect{s_\iZ^1}\) and \(\vect{s_\iZ^2}\).}
    \label{fig:example}
\end{figure}

We demonstrate SOGRAND decoding as illustrated in Fig.~\ref{fig:overview} using a small-sized generalized LDPC code, depicted in Fig.~\ref{fig:example}. Although too small to be a quantum Tanner code, it has all of the relevant properties that \ac{SOGRAND} decoding exploits for quantum Tanner codes. The code protects 5 qubits using two constraint nodes, each with a component code defined by \(H_\mathrm{Xc}\), resulting in the global parity-check matrix \(H_\iX\) with block structure. We further assume that the qubits experience the phase error \(\vect{e_\iZ}\), resulting in the local syndromes \(\vect{s_\iZ^1}\) and \(\vect{s_\iZ^2}\).

Assume that during decoding, the following \acp{LLR} \(\vect{L_\mathrm{A}^1} = (-0.1, -0.9, 1.5)^T\) are sent to constraint node \(1\). These \acp{LLR} represent the beliefs that each bit in \ac{CN} 1's local view is \(1\), with probabilities \((52, 71, 18) \%\) . At the \ac{CN}, ORBGRAND computes the hard-decision \(\vect{e_\mathrm{hd}} = (1, 1, 0)^T\). In this example, before decoding, \(\vect{e_\mathrm{hd}}\) does not match the local view of the correct error pattern. ORBGRAND carries out the following guesswork for a list size \(\mathcal{L}_\text{max} = 1\):
\begin{align*}
\vect{z}^{(1)} = \begin{pmatrix} 0 & 0 & 0 \end{pmatrix}^T &\rightarrow H_\mathrm{Hc} (\vect{e_\mathrm{hd}} \oplus \vect{z}^{(1)}) = \begin{pmatrix} 0 & 1 \end{pmatrix}^T, \\
\vect{z}^{(2)} = \begin{pmatrix} 1 & 0 & 0 \end{pmatrix}^T &\rightarrow H_\mathrm{Hc} (\vect{e_\mathrm{hd}} \oplus \vect{z}^{(2)}) = \begin{pmatrix} 1 & 1\end{pmatrix}^T,  \\
\vect{z}^{(3)} = \begin{pmatrix} 0 & 1 & 0 \end{pmatrix}^T &\rightarrow H_\mathrm{Hc} (\vect{e_\mathrm{hd}} \oplus \vect{z}^{(3)}) = \begin{pmatrix} 1 & 0\end{pmatrix}^T \checkmark.
\end{align*}
The last guess yields an error pattern consistent with the local syndrome \(\vect{s^1_\iZ}\): \(\mathcal{L} = \{\vect{w}^{(3)} = \vect{z^{(3)}} \oplus \vect{e_\mathrm{hd}}\} = \{(1, 0, 0)^T\}\). SOGRAND computes from \(\mathcal{L}\) the probability mass of the explored guesses \(P_\text{g} = 0.7\) and \(\vect{L_\mathrm{APP}^1} = (-0.3, 1.4, 1.4)\). These LLRs indicate that, after component decoding, the belief that each bit in the \ac{CN}'s local view is 1 is now \((57, 19, 19) \%\), respectively. These beliefs are consistent with the local view of the correct error pattern: through component decoding, the decoder identified the correct error pattern locally. In the next iteration, this knowledge is shared with the connected \ac{CN} via message passing.

\subsection{Decoding Performance}
We evaluate the performance of \ac{SOGRAND} decoding on the quantum Tanner codes $[[200, 10, 10]]$ and $[[250, 10, 15]]$ introduced in Ref.~\cite{radeboldExplicitInstancesQuantum2025}, both of which have max stabilizer weight \(12\). Unlike Ref.~\cite{radeboldExplicitInstancesQuantum2025}, which flattens the Tanner construction into a global parity-check matrix for standard \ac{BP}, we extract the underlying Tanner code graphs and the \((25, 19)\) component codes, and decode the Tanner codes as \ac{GLDPC} codes directly with \ac{SOGRAND}. 
Our implementation and the numerical results are publicly available~\cite{quantumTannerSOGRANDCode}.
We evaluate the decoder on a depolarizing channel with total physical error rate \(p\), where each qubit undergoes an \(X\), \(Y\), or \(Z\) error, each with probability \(p/3\). For independent \(X\) and \(Z\) decoding, we map the depolarizing channel to two independent binary symmetric channels with effective flip probability \(\tilde{p} = 2/3 p\) and set the initial \ac{LLR} to \(L_{\mathrm{ch},i} = \log((1-\tilde{p}) / \tilde{p})\). For the correlation aware variant, we initialize the beliefs to \(P_{\mathrm{ch},i}^\mu = (1-p)\) for \(\mu = I\) and \(P_{\mathrm{ch},i}^\mu = p/3\) otherwise.

\begin{figure*}
    \centering
    \subfloat[{\([[250, 10, 15]]\)} Quantum Tanner Code]{\includegraphics{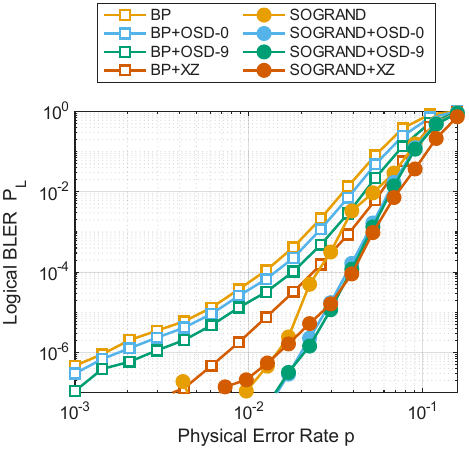}}
    \hfill
    \subfloat[{\([[200, 10, 10]]\)} Quantum Tanner Code]{\includegraphics{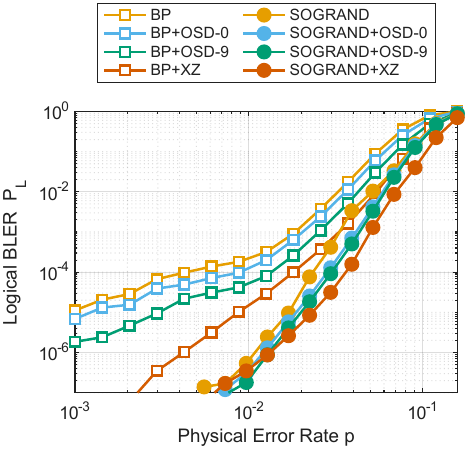}}
    \caption{Quantum Tanner code decoding for BP and SOGRAND decoding with and without additional enhancements. All decoders use a maximum of \(N_\text{iter}=200\) decoding iterations. BP uses the sum-product algorithm with normalized message passing. OSD-\(w\) postprocessing uses the combination sweep strategy with order \(w\).}
    \label{fig:main-results}
\end{figure*}

Figure~\ref{fig:main-results} summarizes the decoding performance (see ``BP'' and ``SOGRAND'' curves). As a baseline, we evaluate BP with the sum-product algorithm and message normalization~\cite{yazdaniImprovingBeliefPropagation2004, laiLogDomainDecodingQuantum2021a}  using the LDPCv2 package~\cite{Roffe_LDPC_Python_tools_2022, roffe_decoding_2020}
\footnote{We adjusted the package to support message normalization for the sum-product algorithm.}. For both BP and SOGRAND, we use a maximum of \(N_\mathrm{iter} = 200\) decoding iterations.
For standard \ac{BP}, to mitigate cycles, message normalization by \(\alpha > 0\) scales check-to-variable-node messages \(L_{i \leftarrow j}^{c}\) to \(L_{i \leftarrow j}^{c} / \alpha\) before passing them to the \acp{VN}. We determined optimal scaling parameters for message normalization via a sweep at a physical error rate of \(p = 10^{-2}\) yielding \(1/\alpha= 0.8\) and \(1/\alpha= 0.85\) for the \([[250, 10, 15]]\) and \([[200, 10, 10]]\) quantum Tanner code, respectively. For SOGRAND, message normalization is not required for these quantum Tanner code instances since it typically yields only negligible performance gains.
SOGRAND outperforms BP with message normalization by more than \(2\) orders of magnitude and achieves higher pseudo thresholds (see Sec.~\ref{sec:conv} and Fig.~\ref{fig:conv-pseudo-thre}).

\subsection{Convergence Behavior}\label{sec:conv}
\begin{figure*}
    \centering
    \subfloat[Logical BLER at \(p = 10^{-1.8}\) \label{fig:conv-bler}]{\includegraphics{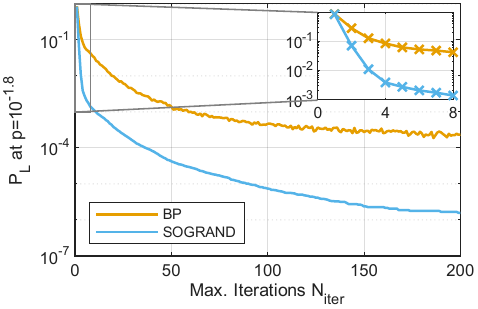}}
    \hfill
    \subfloat[Pseudo threshold \label{fig:conv-pseudo-thre}]{\includegraphics{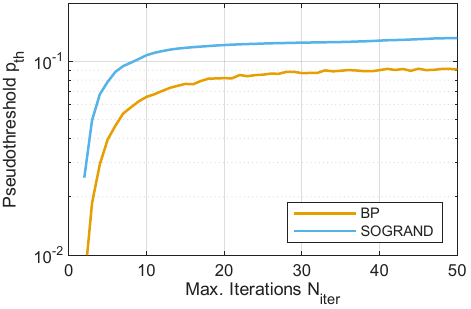}}
    \caption{Convergence behavior of logical BLER and pseudo threshold as a function of the maximum decoding iterations \(N_\text{iter}\) of the {\([[250, 10, 15]]\)} quantum Tanner code.}
    \label{fig:convergence}
\end{figure*}

The superior performance of SOGRAND is achieved with remarkably fast convergence compared to \ac{BP}. Figure~\ref{fig:conv-bler} quantifies this convergence speed and shows the logical \ac{BLER} at \(p=10^{-1.8}\) for both \ac{BP} (sum-product decoding with optimized message normalization
\footnote{For the \([[200, 10, 10]]\) code at \(p=10^{-3}\), min-sum with \(1/\alpha = 0.625\) performs slightly better, otherwise, the normalized sum-product algorithm yields better results.}) and \ac{SOGRAND} as a function of \(N_\mathrm{iter}\). Due to the stronger component codes, \ac{SOGRAND} realizes low logical error rates much faster than \ac{BP}, achieving after \(10\) iterations the same logical \ac{BLER} that \ac{BP} requires more than \(50\) iterations to attain. Furthermore, while \ac{BP}'s decoding performance starts to saturate after \(100\) iterations, SOGRAND continues to gain performance, reducing logical BLER by nearly one decade from \(100\) to \(200\) iterations and another decade from \(200\) to \(3620\) iterations (see Fig.~\ref{fig:relay-results}).
This advantage is reflected in the pseudo-threshold \(p_\mathrm{th}\) (see Fig.~\ref{fig:conv-pseudo-thre}), defined as the physical error rate at which the logical \ac{BLER} of the \(n\) coded qubits equals that of \(k\) uncoded qubits, \(1-(1-p)^k\). For small \(p\), the \ac{BLER} of \(k\) uncoded qubits is approximately \(k\) times the physical error rate, \(k p\).

The primary bottleneck for real-time quantum error correction is algorithmic \emph{depth}, i.e., the sequence of operations that must be performed serially owing to data dependencies. 
Because decoding iterations are inherently sequential, the total latency is the product of the iteration count and the depth per iteration. For \ac{SOGRAND} decoding, the internal querying process is inherently parallelizable~\cite{duffyOrderedReliabilityBits2022}. For the analyzed codes, which have \(6\) redundancy bits per component code, each candidate error pattern in \(\mathcal{L}\) is typically found in fewer than \(2^6 = 64\) parallel queries. Because the local component codes have a fixed length~\cite{leverrierQuantumTannerCodes2022}, the algorithmic depth per iteration is independent of the codelength \(n\). By combining strong BLER performance at a low iteration count with the constant iteration depth, \ac{SOGRAND}-based decoding offers a scalable framework for low-latency quantum error correction.

\section{Improvements}\label{sec:impr}
Since SOGRAND decoding relies on iterative message passing, many existing enhancements for standard \ac{BP} can be adopted, including neural belief propagation~\cite{liuNeuralBeliefPropagationDecoders2019a, miaoQuaternaryNeuralBelief2025add} and quaternary \ac{BP}~\cite{laiLogDomainDecodingQuantum2021a}. In particular, since \ac{SOGRAND} produces soft-output, postprocessing techniques~\cite{panteleevDegenerateQuantumLDPC2021} and guided decimation~\cite{yaoBeliefPropagationDecoding2024aadd} are compatible. Since standard \ac{BP} and \ac{SOGRAND}-based \ac{GLDPC} decoding use the same \ac{VN} updates, memory effect techniques~\cite{kuoExploitingDegeneracyBelief2022add, chenImprovedBeliefPropagation2025add, chytasEnhancedMinSumDecoding2025a, mullerImprovedBeliefPropagation2025} are compatible as well.
We therefore consider component decoding with \ac{SOGRAND} as a new base decoder that can replace standard \ac{BP} in many of these modifications if the code admits a suitable \ac{GLDPC} structure. In this section, we apply three such techniques to \ac{SOGRAND} and compare against \ac{BP} using the same enhancements.

\subsection{OSD Postprocessing}
To mitigate convergence issues inherent to \ac{QLDPC} codes, \ac{BP}+\acs*{OSD}~\cite{panteleevDegenerateQuantumLDPC2021} uses \ac{OSD}~\cite{fossorierSoftdecisionDecodingLinear1995add} as a postprocessing step if \ac{BP} fails to converge. Should \ac{BP} fail, \ac{OSD}-\(w\) reorders bits by decreasing reliability, derived from the \ac{BP} soft-output, and uses the most reliable information set---found during reliability sorting---to generate the unreliable bit set via Gaussian elimination on \(H_\iX\). The decoder then explores \(2^w\) candidate patterns by systematically flipping the \(w\) least reliable bits in the information set. For each flipping, the bits outside of the information set are extended to satisfy the syndrome condition, and the pattern with the highest likelihood is selected. The complexity of \ac{OSD}-\(w\) postprocessing is in general $O(n^3+n^{w+1})$.

Since \ac{OSD} operates on the soft-output of message-passing, it can be applied to \ac{SOGRAND} decoding. Fig.~\ref{fig:main-results} compares BP and SOGRAND with and without OSD postprocessing (``BP+OSD'' and ``SOGRAND+OSD'' versus ``BP'' and ``SOGRAND'' curves). Here, BP+OSD uses the same message-normalization scaling parameters as for \ac{BP} without OSD.
We evaluate BP+OSD and SOGRAND+OSD using both \ac{OSD}-\(0\) and \ac{OSD}-\(9\), each implemented using the LDPCv2 package with combination sweep strategy~\cite{Roffe_LDPC_Python_tools_2022}.
For both quantum Tanner codes considered, OSD postprocessing achieves comparable performance gains for BP and SOGRAND, and the performance gap between BP-based and SOGRAND-based decoding persists. This demonstrates that even with \ac{OSD} postprocessing, \ac{BP} provides suboptimal performance on these quantum Tanner code instances. Interestingly, for \ac{SOGRAND}, OSD order 0 is sufficient to realize most of the OSD gains.

\subsection{Correlation-aware decoding}\label{sec:corr-aware-dec}

\begin{figure}
    \centering
    \includegraphics{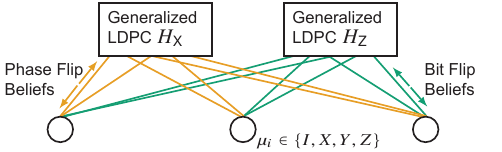}
    \caption{Message passing for correlation-aware SOGRAND decoding. Variable nodes represent Pauli operators \(I, X, Z, Y\) rather than bits. Inspired by Lai and Kuo~\cite{laiLogDomainDecodingQuantum2021a}, scalar messages, here in the form of bit and phase flip beliefs, are passed to a generalized LDPC code of the X and Z stabilizers, respectively.}
    \label{fig:xz-message-passing}
\end{figure}

For \ac{QLDPC} codes, \(X/Z\) correlation can be incorporated via non-binary \ac{BP}. While initial proposals with quaternary message passing in~\cite{babarFifteenYearsQuantum2015a} only apply to linear codes, Lai and Kuo introduced an efficient scalar message-passing scheme~\cite{laiLogDomainDecodingQuantum2021a} that generalizes to additive $\mathrm{GF}(4)$ codes---of which quantum codes are a special case.
Since quantum Tanner codes are \ac{CSS} codes with independent \(X\) and \(Z\) stabilizers, \(X/Z\) correlation can also be integrated into the \ac{GLDPC} framework with minimal overhead by decoding \(X\) and \(Z\) errors simultaneously on their respective classical Tanner codes. This decoder variation can be interpreted as a generalization of Ref.~\cite{laiLogDomainDecodingQuantum2021a} to multi-constraint component codes, applied to \ac{CSS} codes. 

In this case, as shown in Fig.~\ref{fig:xz-message-passing}, each variable node \(i \in [n]\) is connected to two \acp{CN} from each Tanner code, allowing the fusion of incoming beliefs from both subgraphs. To fuse beliefs, we transform each incoming \ac{LLR}, \(L_{i \leftarrow j}^{\mathrm{c}}\), into beliefs \(P_{i \leftarrow j}^{\mathrm{c},\mu}\) for all four Pauli operators \(\mu \in \mathcal{P} = \{I, X, Y, Z\}\). This is done by accounting for the fact that messages from \(H_\iZ\) and \(H_\iX\) carry no information about \(Z\)- or \(X\)-errors, respectively. For example, for an incoming message from \(H_Z\), the mapping is:
\(
    P_{i \leftarrow j}^{\mathrm{c},\iX} = P^{\mathrm{c},\iY}_{i \leftarrow j} = 0.5 q
\) and
\(
    P_{i \leftarrow j}^{\mathrm{c},\iI} = P^{\mathrm{c},\iZ}_{i \leftarrow j} = 0.5 (1 - q),
\)
where \(q = (1 + \exp(L_{i \leftarrow j}^{\mathrm{c}}))^{-1}\) is the corresponding bit flip probability. 
\acp{APP} are obtained by normalizing the product of the channel prior \(\vect{P_\mathrm{ch}^\mu}\) and all incoming symbol beliefs: 
\[
    P_{\mathrm{APP},i}^\mu \propto P_{\mathrm{ch},i}^\mu \prod_{j \in \mathcal{N}(i)} P_{i \leftarrow j}^{\mathrm{c},\mu}.
\]
Extrinsic messages are computed by excluding the respective incoming message from the product: \(P_{i \rightarrow j}^{\mathrm{v},\mu} \propto P_{\mathrm{APP},i}^\mu / P_{i \leftarrow j}^{\mathrm{c}, \mu}\). Finally, these are marginalized back to bit and phase-flip \acp{LLR} messages and passed to the \acp{CN}.

We note that, as the correlation-aware variant can be interpreted as a generalization of Ref.~\cite{laiLogDomainDecodingQuantum2021a}, the proposed framework can be extended to other quantum codes with local component codes, including non-\ac{CSS} codes, as long as, for each component code, all entries within each column of its check matrix are identical. Assuming such a component code \(\tilde{H}_\mathrm{c} \in \mathcal{P}^{m_\mathrm{c} \times n_\mathrm{c}}\), Eq.~\eqref{eqn:syndrome-condition2} becomes 
\(
    s^j_t = \bigoplus_{i = 1}^{n_\mathrm{c}} \tilde{H}_{\mathrm{c},t,i} \ast \tilde{W}_i,
\)
for \(t \in [m_\mathrm{c}]\) and \(\vect{s^j} \in \fbinary^{m_\mathrm{c}}\).
Here, \(\tilde{W} \in \mathcal{P}^{n_\mathrm{c}}\) is a general error pattern and \(\tilde{H}_{t,i} \ast \tilde{W}_i\) is the symplectic product, which is \(0\) if \(\tilde{H}_{t,i}\) and \(\tilde{W}_i\) commute, and \(1\) otherwise. Analogous to Eq.~\eqref{eqn:syndrome-condition2}, this equation represents a sum of binary variables. A generalized version of the proposed decoder simply passes beliefs about \(\tilde{H}_{t,i} \ast \tilde{W}_i\) along the edges~\cite{laiLogDomainDecodingQuantum2021a}, decoding binary component codes, which leaves \ac{SOGRAND} component decoding completely unchanged.

Figure~\ref{fig:main-results} compares the SOGRAND with correlation-aware decoding (``SOGRAND+XZ'' curve) with quaternary BP~\cite{laiLogDomainDecodingQuantum2021a} (``BP+XZ'' curve) for the depolarizing channel, where all decoders use \(N_\text{iter} = 200\) decoding iterations. As for the BP baseline, we use message normalization for BP+XZ. The normalization factors are optimized at \(p = 10^{-2}\) and are \(1/\alpha=0.85\) and \(1/\alpha=0.925\), for the \([[250, 10, 15]]\) and \([[200, 10, 10]]\) quantum Tanner code, respectively. For simulation of BP+XZ, we use a modified version of \footnote{Code: \url{https://github.com/kit-cel/quantum-neural-bp4-demo}} introduced in Ref.~\cite{miaoQuaternaryNeuralBelief2025add}.
The results demonstrate that SOGRAND+XZ can efficiently exploit noise correlation, further boosting performance over the standard SOGRAND decoding at moderate physical error. This enhanced performance comes with minimal additional computational overhead.

In summary, \ac{SOGRAND} performs close to \ac{SOGRAND}+OSD and \ac{SOGRAND}+XZ at low and high physical error rates, while at intermediate error rates the secondary modifications give rise to additional gains. 
In particular, \ac{SOGRAND}+XZ performs best across a wide range of physical error rates, with no need for \ac{OSD}, which may be sufficient for many applications.
Sec.~\ref{sec:llr-init} attributes \ac{SOGRAND}'s performance in the region around \(p=10^{-1.5}\) to the initialization of \(L_{\text{ch},i}\) and explains how this initialization can be improved with a distinct initialization~\cite{miaoQuaternaryNeuralBelief2025add}.

\subsection{Relay Principle}\label{sec:relaybp}
RelayBP~\cite{mullerImprovedBeliefPropagation2025} improves BP performance by introducing memory between iterations. This is realized by introducing iteration-dependent priors, i.e., 
\begin{equation}\label{eqn:channel-calc}
    L_{\mathrm{ch},i}^{(\ell)} = (1-\gamma_i^{(\ell)}) L_{\mathrm{ch},i} + \gamma_i^{(\ell)} L_{\mathrm{APP},i}^{(\ell-1)},
\end{equation}
for iteration \(\ell\), that replace the channel \ac{LLR} \(L_{\mathrm{ch},i}\) in \ac{BP}'s \ac{VN} update. These priors are a weighted average of the channel \ac{LLR} and the posterior \ac{LLR} of the previous iteration \(L_{\mathrm{APP},i}^{(\ell-1)}\). Central to RelayBP are the bit- and iteration-dependent memory strengths \(\gamma_i^{(\ell)} \in \freal\).
Ref.~\cite{mullerImprovedBeliefPropagation2025} achieves impressive performance gains by sampling the memory strengths every \(T_\text{r}\) decoding iterations independently and uniformly from an interval  
\(
    \gamma_i^{(\ell)} \in [\gamma_\text{c} - \gamma_\text{w}/2, \gamma_\text{c} + \gamma_\text{w}/2]
\). Ref.~\cite{mullerImprovedBeliefPropagation2025} refers to such a period of \(T_\text{r}\) iterations as a \emph{leg}. The first leg is an exception with \(T_0\) iterations and a constant memory strength \(\gamma_0\) across bits, resulting in \(T_0 + (R-1) T_\text{r}\) total BP iterations for \(R\) legs. 

As SOGRAND decoding relies on message passing and uses the same \ac{VN} update as standard BP, the Relay principle can be directly integrated into SOGRAND decoding. Specifically, Eq.~\eqref{eqn:vn-app-calc} is replaced by
\[
    L_{\mathrm{APP},i}^{(\ell)} \leftarrow L_{\mathrm{ch},i}^{(\ell)} + L_{i \leftarrow j}^{\mathrm{c},(\ell)} + L_{i \leftarrow j^\prime}^{\mathrm{c},(\ell)},
\]
where \(L_{i \leftarrow j}^{\mathrm{c},(\ell)}, L_{i \leftarrow j^\prime}^{\mathrm{c},(\ell)}\) are the \ac{VN} to \ac{CN} messages in the \(\ell\)-th iteration and \(L_{\mathrm{ch},i}^{(\ell)}\) is obtained at the start of  iteration \(\ell\) via Eq.~\eqref{eqn:channel-calc}.

\begin{figure}
    \centering
    \includegraphics{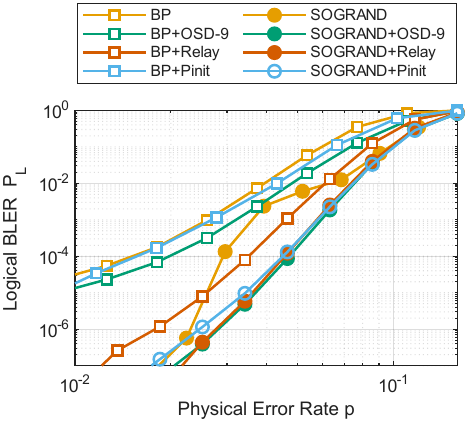}
    \caption{Decoding of the \([[250, 10, 15]]\) quantum Tanner code with the Relay principle with \(R=60\) (\(N_\text{iter} = 3620\)) decoding iterations. For reference, standard BP and SOGRAND decoding, with and without OSD postprocessing, are also shown with \(N_\text{iter} = 3620\) decoding iterations. BP+Pinit and SOGRAND+Pinit initialize \(L_{\text{ch},i}\) with \(p_\text{init} = 10^{-1.8}\).}
    \label{fig:relay-results}
\end{figure}

We numerically evaluate the decoding performance of RelayBP and SOGRAND decoding with the Relay principle, here referred to as BP+Relay and SOGRAND+Relay for the \([[250, 10, 15]]\) quantum Tanner code
\footnote{BP+Relay results are generated via the official RelayBP implementation of Ref.~\cite{mullerImprovedBeliefPropagation2025}: \url{https://github.com/trmue/relay}}.
Both BP+Relay and SOGRAND+Relay use \(R= 60\) legs and, following Ref.~\cite{mullerImprovedBeliefPropagation2025}, \(T_0 = 80\) and \(T_\text{r} = 60\) decoding iterations per leg, resulting in \(3620\) total decoding iterations. Since RelayBP's performance can be significantly improved by an appropriate choice of memory strengths~\cite{mullerImprovedBeliefPropagation2025}, we optimize \(\gamma, \gamma_\text{c}\) and \(\gamma_\text{w}\) for both decoders, as outlined in Appendix~\ref{sec:relay-hyper-parameter}. The optimized parameters are listed in Tab.~\ref{tab:opt-parameters}.
\begin{table}[tbp]
\caption{\label{tab:opt-parameters} Optimal parameters for BP+Relay and SOGRAND+Relay for the \([[250, 10, 15]]\) quantum Tanner code}
\begin{ruledtabular}
\begin{tabular}{lccc}
  & \(\gamma\) &  \( \gamma_\text{c}\) & \(\gamma_\text{w}\) \\
  \colrule
  BP+Relay & \(0.07\) & \(0.18\) & \(0.64\)\\
  SOGRAND+Relay &  \(0.07\) & \(-0.15\) & \(0.52\) \\
\end{tabular}
\end{ruledtabular}
\end{table}

\begin{figure}[tbp]
    \centering
    \includegraphics{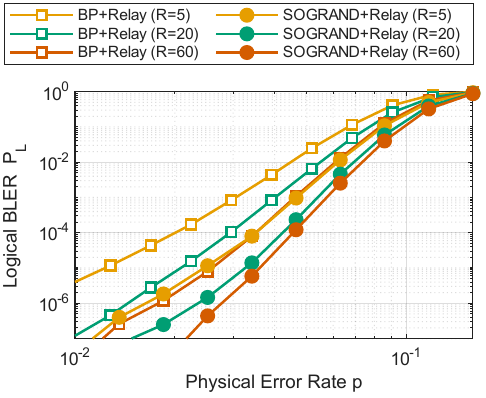}
    \caption{BP+Relay and SOGRAND+Relay decoding performance for relay leg counts \(R \in \{5, 20, 60\}\). The decoders are evaluated for the \([[250, 10, 15]]\) quantum Tanner code and fixed iterations per leg: \(T_0 = 80\) and \(T_\text{r} = 60\). SOGRAND+Relay achieves, with \(R=5\), the decoding performance of BP+Relay with \(R=60\).}
    \label{fig:set-sweep}
\end{figure}

Figure~\ref{fig:relay-results} shows the decoding performance for the \([[250, 10, 15]]\) quantum Tanner code using BP+Relay and SOGRAND+Relay with \(R=60\) (\(N_\text{iter}=3620\)). Also included for reference is the decoding performance of BP and SOGRAND with and without OSD postprocessing, also with \(3620\) decoding iterations. SOGRAND+Relay outperforms BP+Relay by around one decade in logical \ac{BLER} at low physical error rates. SOGRAND+Relay matches SOGRAND+OSD with \(3620\) decoding iterations without requiring complex OSD postprocessing. While we choose a high number of decoding iterations to show saturated decoding performance, SOGRAND+Relay also achieves excellent performance with reduced iteration counts. To demonstrate this, we evaluate in Fig.~\ref{fig:set-sweep}, BP+Relay and SOGRAND+Relay's performance for leg counts \(R \in \{5, 20, 60\}\): SOGRAND+Relay achieves, with \(R=5\) legs, Relay+BP's performance with \(R=60\), requiring only \(320\) decoding iterations compared to \(3620\).

\subsection{Channel LLR Initialization}\label{sec:llr-init}
Following standard practice~\cite{roffe_decoding_2020,mullerImprovedBeliefPropagation2025}, in the previous sections, every \ac{BP} and \ac{SOGRAND} variant initializes the statistical input \(L_{\text{ch},i}\) and \(P_{\text{ch},i}^\mu\) based on the physical error rate \(p_\text{init} = p\). 
For QLDPC codes, Miao et al. observed additional gains with a fixed initialization \(p_\text{init}\), independent of \(p\)~\cite{miaoQuaternaryNeuralBelief2025add}. As shown in Fig.~\ref{fig:relay-results} (SOGRAND+Pinit), for the quantum Tanner code with \ac{SOGRAND} decoding, such a fixed initialization with \(p_\text{init} = 10^{-1.8}\) provides additional gains, outperforming BP+Relay across \(p\). For \ac{BP}, mildly enhanced performance can be obtained with fixed initialization.

Iterative SISO decoding works by separately looking at sub-components of the code in isolation and then sharing updated beliefs with neighboring components, ultimately converging to a global solution. In classical communications applications, the starting beliefs from the channel are nuanced with distinct per-bit reliabilities. In quantum error correction, the starting beliefs are statistical in nature. By initializing the decoder with more conservative beliefs based on \(p_\text{init} = 10^{-1.8}\), lower-Hamming-weight corrections are given greater emphasis in local updates, which results in a more restrained belief propagation in early iterations and improved performance.

\section{Conclusion}
This work proposes \ac{SOGRAND} decoding for quantum Tanner codes.
Performance evaluation on explicit instances demonstrates that the proposed decoder outperforms the established BP+OSD baseline by over two orders of magnitude in logical \ac{BLER}. 
This effectively removes the cubic complexity bottleneck associated with \ac{OSD} and highlights the great potential of the quantum Tanner code class, as it admits a far more effective decoder. Since SOGRAND relies on iterative message passing, existing optimization techniques for standard \ac{BP} can be readily adopted. We demonstrate this with OSD postprocessing, correlation-aware decoding, and the Relay principle, each yielding additional performance gains over the baseline approach. Combined with the Relay principle, SOGRAND outperforms RelayBP by over an order of magnitude in logical error rate.
Although presented within the code capacity setting~\cite{dennisTopologicalQuantumMemory2002}, 
none of the presented techniques inherently rely on error-free syndrome measurements. Following similar extensions for \ac{BP}~\cite{kuoGeneralizedQuantumDataSyndrome2025}, the noise model can be relaxed to phenomenological and circuit-level noise~\cite{dennisTopologicalQuantumMemory2002, stephensFaulttolerantThresholdsQuantum2014}. Characterizing \ac{SOGRAND}'s performance under these models is a promising direction for future work. These results suggest a viable path toward practical, scalable implementations of quantum Tanner codes.

\section{Note} During the completion of this manuscript, we became aware of concurrent research on the iterative decoding of quantum Tanner codes~\cite{mostad2026}. The primary distinction between these works is that Ref.~\cite{mostad2026} focuses broadly on decoding quantum codes with generalized LDPC structure, with quantum Tanner codes as a paradigmatic example, whereas our work centers on the use of SOGRAND as a highly effective SISO decoder. In Ref.~\cite{mostad2026}, component code decoding is performed using an exact MAP decoder implemented via the BCJR algorithm. This provides an upper bound on the decoding performance, but is computationally inefficient. Indeed, Ref.~\cite{mostad2026} identifies the development of an efficient SISO decoding algorithm for the component codes as an open problem, which is resolved here using SOGRAND.\\

\section*{Data Availability}
The data that support the findings of this article are openly available~\cite{quantumTannerSOGRANDCode}.

\appendix
\section{RelayBP Hyperparameter Optimization \label{sec:relay-hyper-parameter}}

\begin{figure*}[tbp]
    \centering
    \subfloat[Sweep $\gamma$ first leg \label{fig:gamma0-sweep}]{\includegraphics{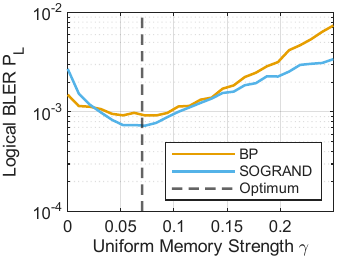}}
    \hfill
    \subfloat[Interval memory strength sweep (BP) \label{fig:sweep-bp}]{\includegraphics{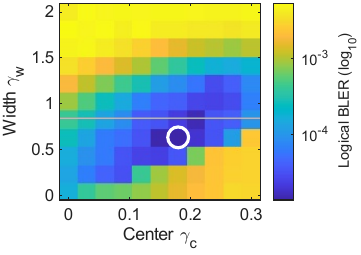}}
    \hfill
    \subfloat[Interval memory strength sweep (SOGRAND) \label{fig:sweep-sogrand}]{\includegraphics{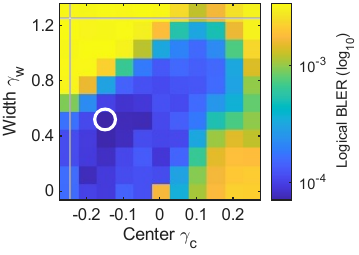}}

    \caption{Hyperparameter optimization for the Relay principle on the {{\([[250, 10, 15]]\)}} quantum Tanner code. BP+Relay and SOGRAND+Relay: the vertical dashed line and the white circles mark the optimal hyperparameter sets.}
    \label{fig:hyperparameter-opt}
\end{figure*}

We perform hyperparameter optimization for BP+Relay and SOGRAND+Relay on the \([[250, 10, 15]]\) quantum Tanner code, following the approach from~\cite{mullerImprovedBeliefPropagation2025}~\footnote{Ref.~\cite{mullerImprovedBeliefPropagation2025} does not specify how the uniform memory strength for the first leg is determined.}. The optimization consists of a uniform memory-strength (\(\gamma\)) sweep and an interval (\(\gamma_\text{c},\gamma_\text{w}\)) sweep, as shown in Fig.~\ref{fig:hyperparameter-opt}.
The uniform memory strength sweep operates around a target logical error rate of \(10^{-3}\) in the waterfall region (\(p = 10^{-1.45}\) for SOGRAND+Relay and \(p = 10^{-1.65}\) for BP+Relay). Interval sweeps use slightly higher physical error rates to reduce simulation variance (\(p = 10^{-1.35}\), and \(p = 10^{-1.55}\), respectively).

To find the optimal uniform memory strength \(\gamma\), we terminate after \(T_0 = 80\), and sweep over \(\gamma\) as visualized in Fig.~\ref{fig:gamma0-sweep}, resulting in an optimal \(\gamma\) of \(0.07\) for both BP+Relay and SOGRAND+Relay. For fixed \(\gamma=0.07\), we run BP+Relay and SOGRAND+Relay for the \(R=60\) legs and vary the width \( \gamma_\text{w}\) and center \(\gamma_\text{c}\) of the sampling interval as shown in Fig.~\ref{fig:sweep-bp} and \ref{fig:sweep-sogrand}. 

Ref.~\cite{mullerImprovedBeliefPropagation2025} already observed that the Relay principle requires negative memory strengths for good performance. For SOGRAND decoding with the Relay principle, this effect is even more pronounced. The optimal sampling interval for BP+Relay has \(22 \%\) negative memory strengths, comparable to the optimal intervals in Ref.~\cite{mullerImprovedBeliefPropagation2025} for Gross and surface codes, whereas for SOGRAND the optimal interval has \(79 \%\) negative memory strengths.

\end{document}